\documentclass[12pt]{article}
\usepackage{graphicx}
\begin{document}
\begin{center}
{\large \bf On the amplification of diffusion on piecewise linear
potentials by direct current}
\end{center}
\begin{center}
Els Heinsalu, Risto Tammelo\footnote{Corresponding author. Fax.
+372-7-375570; Tel. +372-7-375571; E-mail:
\mbox{tammelo@ut.ee}}\hspace{5pt} and Teet \"{O}rd
\\ \textit{Institute of Theoretical Physics, University of Tartu,\\T\"{a}he
4, 51010 Tartu, Estonia}
\end{center}
\vspace{0.5 cm}
\begin{abstract}
The diffusive motion of overdamped Brownian particles in tilted
piecewise linear potentials is considered. It is shown that the
enhancement of diffusion coefficient by an external static force
is quite sensitive to the symmetry of periodic potential. Another
new effect found is that the factor of randomness as a function of
the tilting force exhibits a plateau-like behaviour in the region
of low temperatures. \vspace{1 cm}

\end{abstract}
\pagenumbering{arabic} \vspace{1cm} Various aspects of the
dynamics of Brownian particles in external periodic potentials
have been a subject of growing interest in the last few years
[1-3]. The interplay of noise and nonlinear dynamics produces a
rich variety of remarkable physical effects, for example, the
dependence of the (direction of) particle current on the
statistical properties of the nonequilibrium fluctuations which
induce the current [1,4-6]. The present paper addresses the
interrelationship between the current and diffusion of the
Brownian particles on tilted simple sawtooth potentials. In the
case of thermal equilibrium, the overdamped force-free thermal
diffusion of mutually non-interacting Brownian particles is always
reduced when a additional periodic potential is applied to the
system, as corresponding localization of the particles occurs
\cite{lifson}. Therefore one tends to think that a qualitatively
similar behaviour is valid at least for some time-independent
non-equilibrium systems. However, recently it was discovered that
the opposite is the case: in the seminal papers \cite{5,6} it was
demonstrated that a giant enhancement of diffusion will take place
if, in addition to a periodic potential, a constant tilting force
is  applied. Refs. \cite{5,6} predict that the enhancement of
diffusion can be as large as even up to fourteen orders of
magnitude under the realistic room temperature conditions. The
rapid increase in the effective diffusion coefficient appears due
to the interaction between the potential and particles moving at a
nonzero mean velocity. This discovery has generated additional
interest in the investigation of the Browninan transport pointing
to new possibilities in modelling experimental data of stochastic
processes in periodic structures. Paper \cite{6} refers to a
number of various applications in this context .

A further development in this direction was to apply to the system
a sinusoidal time-periodic force in addition to the tilted
sinusoidal space-periodic force. It was demonstrated numerically
"that the interplay between the frequency-locking and the noise
gives rise to a multi-enhancement of the effective diffusion, and
a rich behaviour of the current as well, including partial
suppression and characteristic resonances" \cite{7}, see also Ref.
\cite{8} for the effect of the acceleration of diffusion by
time-periodic driving. Mathematically exactly the same overdamped
Langevin equation with the white thermal noise, sinusoidal spatial
potential, sinusoidal time-periodic ac force, and a tilting dc
force as in Ref. \cite{7} was considered in paper \cite{9}, in
which also an analytic solution of the Langevin equation was
found, but the authors focused their attention to the influence of
a strong ac force rather than of the tilting dc force on the
motion of Brownian particles. Another recent development has been
to study the influence of the frictional inhomogeneity of the
medium to the giant diffusion. Paper \cite{10} considered "a
simple minimal model when the potential is sinusoidal and the
friction coefficient is also periodic (sinusoidal) with the same
period, but shifted in phase". It was demonstrated that both the
giant diffusion and the coherence of transport in a tilted
periodic potential are sensitive to the frictional properties of
the medium. The influence of the space dependent friction on the
mobility of an overdamped particle moving in a washboard
potential with bias has been studied in \cite{11}.

Thus, on the basis of the above-mentioned pioneer investigations
one can conclude that in order to obtain a significant
amplification of the diffusion of Brownian particles, the
following minimal ingredients must be present (i) the thermal
noise, (ii) a periodic structure, and (iii) a constant tilting
force. Every additional physical agent may bring forward new
essential physical features, as was already demonstrated by the
inclusion of a force periodic in time \cite{7,8} and a periodic
friction coefficient \cite{10}. In the present communication we
will numerically study the diffusion in a tilted saw-tooth
potential on the basis of the general theoretical scheme developed
in \cite{5,6}. The influence of the asymmetry of the potential on
the behaviour of the diffusion coefficient will be investigated
and the dependencies of the factor of randomness on the tilting
force and temperature will be calculated.

The motion of an overdamped Brownian particle under the influence
of the periodic potential $V_{0}(x)$, static external force $F$
and thermal noise is described by the following Langevin equation
\begin{equation} \label{1}
\eta \dot{x}(t)=-\frac {dV(x)}{dx}+\xi(t),
\end{equation}
\begin{equation} \label{2}
V(x)=V_{0}(x)-Fx,
\end{equation}
where $V(x)$ is the total deterministic potential, $\eta$ is the
friction  coefficient and $\xi(t)$ is the Gaussian white noise
with the mean value $\langle\xi(t)\rangle=0$ and the correlation
function $\langle\xi(t) \xi(t')\rangle=2 \eta k_{B} T
\delta(t-t')$.

The general definition of the effective diffusion coefficient can
be written as
\begin{equation} \label{3}
D:=\lim_{t \to \infty} \frac {\sigma_{x}^{2} (t)}{2 t},
\end{equation}
where the dispersion of co-ordinate reads
\begin{equation} \label{4}
\sigma_{x}^{2} (t):=\langle x^{2}(t) \rangle-\langle x(t) \rangle
^{2}.
\end{equation}

According to Refs. \cite{5,6} the diffusion coefficient for the
model (\ref{1}), with the periodic boundary conditions imposed,
equals ($F\geq0$)
\begin{equation} \label{5}
D=D_{0} \frac { \int \limits_{x_{0}}^{x_{0}+L} I_{\pm}(x) I_{+}(x)
I_{-}(x) \frac {dx}{L}}{ \left [\int \limits_{x_{0}} ^{x_{0}+L}
I_{\pm}(x) \frac {dx}{L} \right ] ^{3}},
\end{equation}
where
\begin{equation} \label{6}
I_{+}(x)=\frac {1}{D_{0}} e^{V(x)/k_{B}T} \int \limits_{x-L}^{x}
e^{-V(y)/k_{B}T}dy,
\end{equation}
\begin{equation} \label{7}
I_{-}(x)=\frac {1}{D_{0}} e^{-V(x)/k_{B}T} \int \limits_{x}^{x+L}
e^{V(y)/k_{B}T}dy.
\end{equation}
Here $L$ is the spatial period of the potential $V_{0}(x)$,
$x_{0}$ is an arbitrary point and $D_{0}=k_{B}T/\eta$ is the
diffusion coefficient if $V_{0}(x)=0$.

The relation between the diffusive and directed components in the
Brownian motion can be characterized by the factor of randomness
\begin{equation} \label{8}
Q:=\frac {2 D}{L \langle \dot{x} \rangle}.
\end{equation}
For the model (\ref{1}) the particle current has the form
\cite{5,6,12}
\begin{equation} \label{9}
\langle \dot{x} \rangle = \frac {1-e^{-LF/k_{B}T}}{\int
\limits_{x_{0}}^{x_{0}+L}I_{\pm} (x) \frac {dx}{L}}.
\end{equation}

From now on we will consider a simple saw-tooth potential $V_{0}$
with the amplitude $A$ and the asymmetry parameter  $k$ ($0<k<L$;
the value $k=L/2$ corresponding to the symmetric potential).

In order to evaluate the integrals in equations
(\ref{5})-(\ref{7}) and (\ref{9}) at $x_{0}=0$, one needs the
expressions of the potential $V_{0}(x)$ in the following regions
\begin{equation} \label{10}
V_{0}(x)=\left\{ \begin{array} {lrrl}A\, \frac {k-x}{k} \: ,& 0\!
& \leq & \!x \leq k\;,\\ A \,\frac {k-x}{k-L} \: ,& k \!& \leq &\! x \leq L\;,\\
A \,\frac {L+k-x}{k} \: ,& L\! & \leq &\! x \leq L+k\;,\\ A\,
\frac {L+k-x}{k-L} \:,\quad &  L+k\! & \leq &\! x \leq 2L\;.
\end{array}\right.
\end{equation}

The large enhancement of the diffusion coefficient as a function
of the external static force $F$ appears below the critical tilt
value $F_{c}$ which is determined as the threshold for $F$ above
which the local minima of $V(x)$ disappear. In the case under
consideration
\begin{equation} \label{11}
F_{c}= \frac {A}{L-k}.
\end{equation}

\begin{figure}[htbp]
\begin{center}
\includegraphics[width=0.6\linewidth]%
{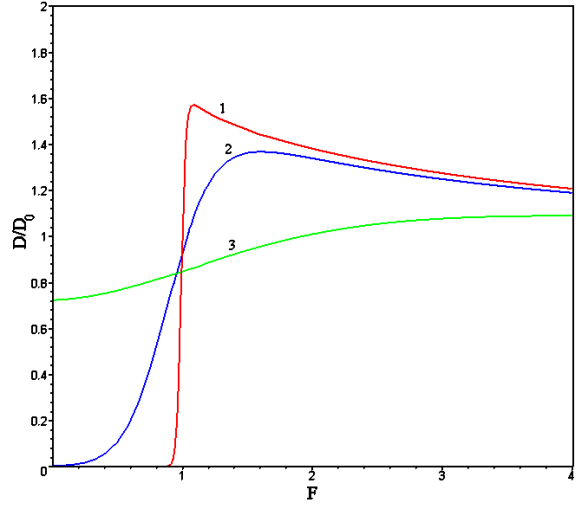} \caption{The diffusion coefficient \textit{vs} the
tilting force
for $k=0.2$. Curve 1 - $T=0.01$, curve 2 - $T=0.1$, curve 3 - $T=0.5$}%
\label{fig:fig1}%
\end{center}
\end{figure}

\begin{figure}[htbp]
\begin{center}
\includegraphics[width=0.6\linewidth]%
{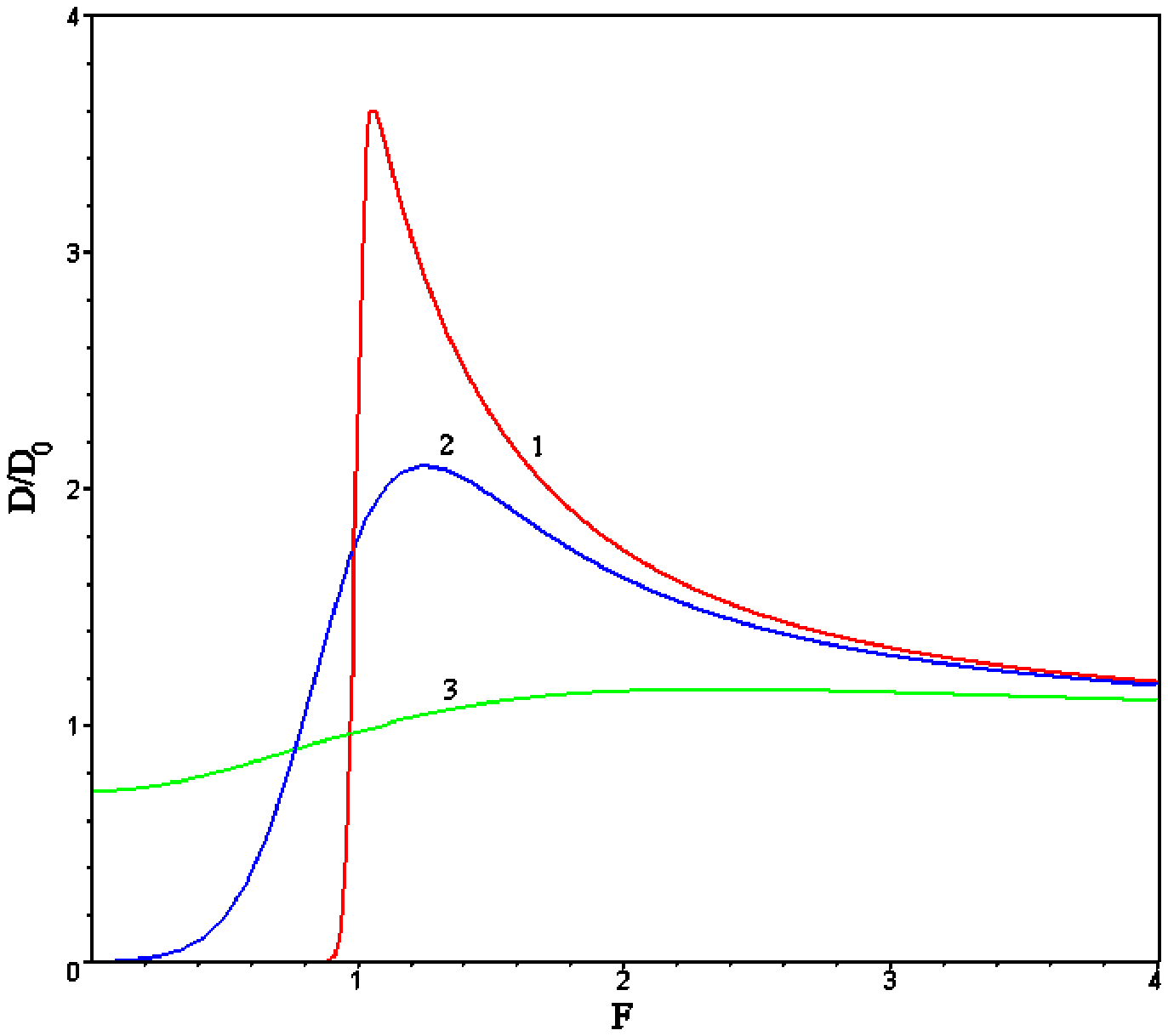}%
\caption{The diffusion coefficient \textit{vs} the tilting force
for $k=0.5$. Curve 1 - $T=0.01$, curve 2 - $T=0.1$, curve 3 - $T=0.5$}%
\label{fig:fig2}%
\end{center}
\end{figure}

\begin{figure}[htbp]
\begin{center}
\includegraphics[width=0.6\linewidth]%
{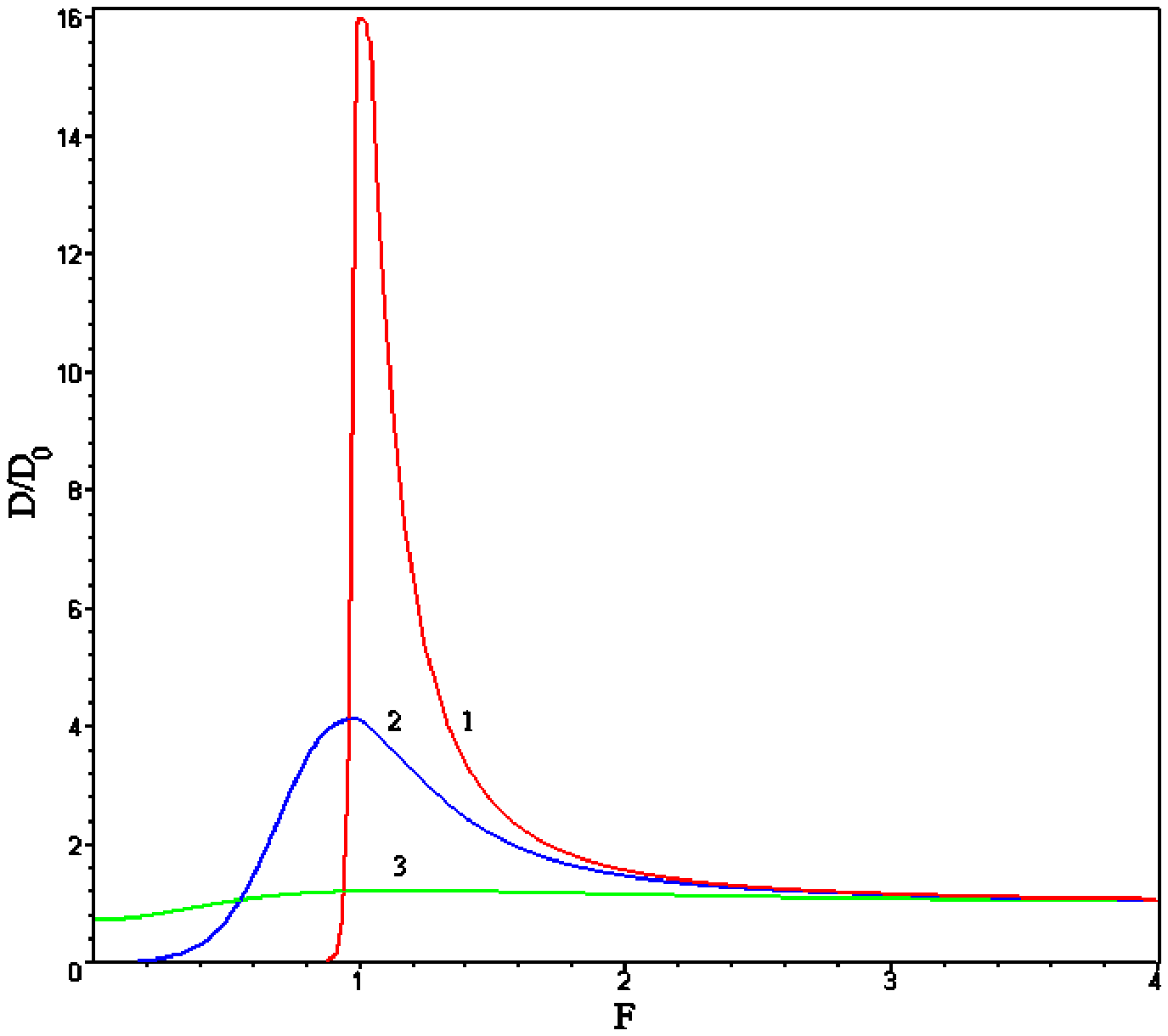}%
\caption{The diffusion coefficient \textit{vs} the tilting force
for $k=0.8$. Curve 1 - $T=0.01$, curve 2 - $T=0.1$, curve 3 - $T=0.5$}%
\label{fig:fig3}%
\end{center}
\end{figure}

In calculations we choose $L=1$ and use the dimensionless
quantities $\tilde{T}:=k_{B}T/A$, $\tilde{F}:=F/F_{c}$,
$\tilde{D}:=\eta D/A$, $\tilde{D_{0}}:=\eta D_{0}/A=\tilde{T}$,
$\tilde{\langle\dot{x}\rangle}:=\eta \langle\dot{x}\rangle/A$,
omitting in what follows the sign tilde for the sake of brevity.

The effective diffusion coefficient $D$ as a function of the
tilting force $F$ at different values of the parameter of
asymmetry $k=0.2$, $0.5$, $0.8$ is shown in Figs. 1-3,
respectively. The function $D(F)$ reveals qualitatively the
analogous behaviour as found in Refs. \cite{5,6}. At the same
time, it is seen from the comparison of Figs. 1-3 that the large
values of $k$ favour the effect of amplification of diffusion with
respect to the free diffusion. In the limit $k\rightarrow1$,
arbitrary large maximal values of $D/D_{0}$ can be obtained, e.g.,
the diffusion coefficient reaches the value $D_{0}\times10^{6}$
at $T=10^{-5}$ and $k=0.999$. However, if the asymmetry parameter
$k \to L$, the critical tilting force $F_{c} \to \infty$, see
equation (\ref{11}).

On the grounds of the above-mentioned properties of the diffusion
and bearing in mind that the tilting force generates the particle
current, one can speculate that there exist two interrelated
channels for increasing $D$. The first one is due to the trivial
delocalization of Brownian particles as a result of tilting: it is
essentially a passive channel. The second channel plays an active
role, leading to a really amplified diffusion, which may exceed
the free diffusion by many orders of magnitude. The crucial factor
here is that there occurs an additional influence of the periodic
potential to the Brownian particles which becomes significant if
the mean velocity of the particles is sufficiently large. In other
words, the character of the interaction between the potential and
current obtains qualitatively new features when the current grows
strong enough. On the other hand, if the current exceeds a certain
critical value determined by the critical tilt, the effect will
disappear.

In Fig. 4 we plot the dependencies of the factor of randomness on
the tilting. The most interesting feature of the curves in Fig. 4
is the presence of a plateau in the wide region of $F$ up to the
critical tilt $F_{c}$ at a sufficiently low temperature.
\begin{figure}[htbp]
\begin{center}
\includegraphics[width=0.6\linewidth]%
{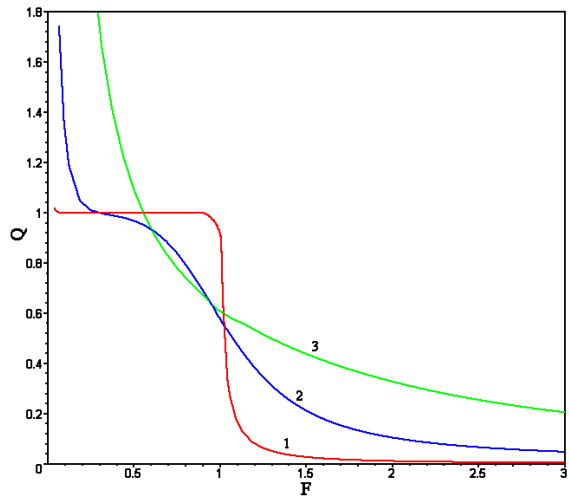}%
\caption{The factor of randomness \textit{vs} the tilting force
for $k=0.5$. Curve 1 - $T=0.01$, curve 2 - $T=0.1$, curve 3 -
$T=0.5$}%
\label{fig:fig4}%
\end{center}
\end{figure}
This means that an extremely neat synchronization of the giant
enhancements of $D$ and $\langle\dot{x}\rangle$ takes place. Let
us emphasize that the plateaus, i.e. the fine tuning of the
directed and diffusive motions occur at the values of system
parameters where the growth rate of the function $D(F)$ is the
largest.
%At the domain of the plateaus one can obtain the
%following analytical expression for the diffusion cofficient $D$
%and current $\langle\dot{x}\rangle$ ($F<1$, $(1-F)/T \gg 1$,
%$F/T(1-k) \gg 1$):
%\begin{equation} \label{12}
%2D= \langle\dot{x}\rangle=
%\frac{(1-F)^{2}(1-(1-F)k)^{2}}{T(1-k)^{2}}e^{(F-1)/T}.
%\end{equation}
At the end of a plateau, $Q(F)$ falls abruptly as $F$ approaches
$F_{c}$. If $F\rightarrow0$, the current turns to zero and
$Q\rightarrow\infty$. With the rise of temperature, the length of
the plateau diminishes from both sides and the fall near $F_{c}$
slows down. Finally the plateau disappears and $Q(F)$ decreases
monotonically. The calculations show that in the region of low
temperatures the variation of the parameter $k$ does not influence
the factor of randomness $Q(F)$. For higher temperatures the
decrease of the function $Q(F)$ becomes more pronounced  if the
asymmetry parameter $k$ increases.

\begin{figure}[htbp]
\begin{center}
\includegraphics[width=0.6\linewidth]%
{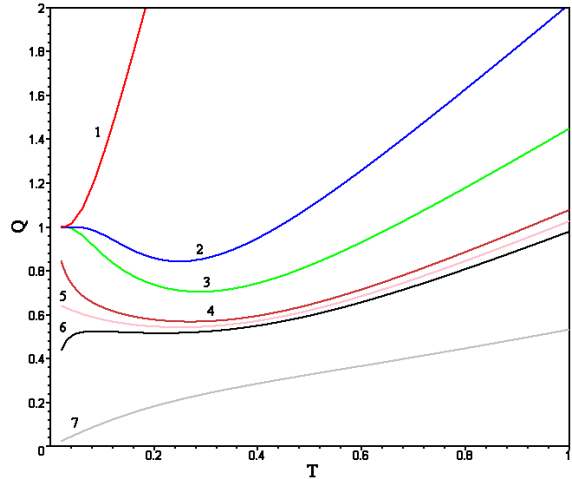}%
\caption{The factor of randomness \textit{vs} temperature for
$k=0.5$. Curve 1 - $F=0.1$, curve 2 - $F=0.5$, curve 3 - $F=0.7$,
curve 4 - $F=0.95$, curve 5 - $F=1$,
curve 6 - $F=1.05$, curve 7 - $F=2$}%
\label{fig:fig5}%
\end{center}
\end{figure}

The behaviour of the factor of randomness %\textit{vs}
as a function of the temperature is displayed  in Fig. 5. It can
be seen that at small (curve 1) or at large (curve 7) values of
the tilting force $F$ the function $Q(T)$ decreases or increases
monotonically. For the intermediate values of $F$ the function
$Q(T)$ passes a minimum which corresponds to a maximum in the
temperature dependence of the P\'{e}clet factor Pe$=2/Q$ (cf.
\cite{10}).

To conclude, our results indicate that the shape  of a periodic
potential (i.e., the asymmetry in our case) essentially influences
the amplification of diffusion by a tilting force. We also
established that there exists a characteristic synchronization
between the diffusion and the current which appears in the region
of parameters where the enhancement of $D(F)$ is most rapid. It
seems that such a stabilization of the coherence level of the
Brownian transport is intrinsically  related to the mechanism of
the amplification effect of diffusion.

\section*{Acknowledgement}

The authors are grateful to Prof. Romi Mankin and Dmitri Martila,
M.Sci., for useful discussions and assistance.

\pagebreak

\end{document}